\documentclass [12pt]{article}

\begin{document}
\title { David Brink : A long-standing teacher}

\author { Angela Bonaccorso\\
\small  Istituto Nazionale di Fisica Nucleare, Sez. di
Pisa, \\
\small and Dipartimento di Fisica, Universit\`a di Pisa,\\
\small Largo Pontecorvo 3, 56127 Pisa, Italy.}
\date{}
\maketitle

\begin{abstract}
This talk presents a short review of David Brink's most important achievements and of my own experience working with him.
 \end{abstract}
 \section{
 David Brink} 
 
 has been, with his ideas and intuitions, one of the founders of Theoretical Nuclear Physics in the way it has been intended in the last fifty years. His thesis at the University of Oxford, 1955, contained the idea of collective motion built on excited states. This effect, confirmed experimentally in the '70s, has been the subject of intense studies both experimentally as well as theoretically. The subject of collective excitations has been a constant presence in David's career.

David has given enormous contributions to the understanding and disentangling of several phenomena and theories of nuclear structure such as the pairing effect or the Generator Coordinate Method and its application to clustering effects. On the nuclear reaction side David has contributed to the developments of microscopic models of the optical potential and he has devoted a long-standing and constant interest in clarifying the mechanism of direct reactions and developing theories for transfer and breakup. His 1972 paper on the transfer matching conditions \cite{1f} remains to date one of the most cited both by theoreticians as well as by experimentalists. During the same period he also developed with D. Vautherin \cite{1e} the famous ideas about the use of effective interactions such as the Skyrme force which are still widely used and which led to the introduction of density functional approaches to the nuclear many-body problem.

David possesses outstanding skills in mathematical computation and he is a great expert on special functions, group theory and the Feynman path integral method. His book with R. Satchler on Angular Momentum \cite{1a} has been studied and will be studied by generations of Physics students. His other books on the Nuclear Force \cite{1b} and on the Semi-classical Methods in Nucleus-Nucleus Scattering \cite{1c} are an unmatched example of scientific clarity, transparency and depth. More recently he has written a book with Ricardo Broglia on Nuclear Super-Fluidity and the Pairing Interaction \cite{1d}. 

David's career is characterized by a large number of students and collaborators from all over the world. For all of them he has been a source of scientific and human inspiration and admiration. His most fundamental teaching has been that research means trying to discover and understand the beauties of Nature and then explain them to the others. His absolute belief in the value of truth and unselfish attitude in sharing knowledge make him an outstanding figure in contemporary Nuclear Physics and one still very active on the front-line of research.

David Brink has been Fellow of the Royal Society since 
1981. In 
1982 he received the Rutherford Medal of the Institute of Physics and in 1992 he was made
 Foreign Member of the Royal Society of Sciences of Uppsala. He has just been awarded  (shared with Prof. 
 Heinz-JŸrgen Kluge) the 2006 Lise Meitner Prize of the European Physical Society.

\section{Our scientific collaboration}
My collaboration with David started in 1978 when I went to Oxford as a graduate student, following the advice of Massimo Di Toro who had been the supervisor of my Master thesis in Catania. During my work in Catania I had used Brink and Boeker \cite{2a} interaction and it was while reading that paper that I was able to understand the origin of a mistake in a code which had retarded my graduation. In this way I  had met David and his guidance capacities even before I actually met him in person.

In Oxford we developed a microscopic model for  the imaginary part of the $\alpha-^{40}$Ca optical potential \cite{1}. Those were the years in which it started to be clear that the imaginary part of the heavy-ion optical potential had not only a volume part but also a surface part due to direct reactions. The Copenhagen group led by A. Winther and R. Broglia in collaboration with  G. Pollarolo  was developing interesting theories on the subject \cite{2}. Thus we also started to look at this problem and we wrote a first simple paper \cite{3} with Giuseppe Piccolo. Giuseppe was my first Master student back in Catania where
I was then working after having finished my D. Phil. He would be also the first third-generation `disciple' of David from my side. That paper was seminal in the use of one-dimensional Fourier transforms of single particle wave functions in a mixed representation. 

In the mean time Luigi Lo Monaco, another student from Catania, moved to Oxford to write a thesis under David's supervision. I spent a few months in Oxford almost every year  and we wrote then a paper on transfer between bound states \cite{4}. That paper contained a lot of wisdom David had accumulated in previous years working with other collaborators such has H. Hasan \cite{5}, Ica Stancu \cite{6,7}, H. Hashim \cite{8}.

Later on David realized that, due to the matching conditions he had discovered and clarified in one of his most cited papers \cite{1f}, transfer would lead preferentially to unbound final states, once that projectiles would  reach  the higher energies ($>$ 10A.MeV) for which accelerators were then under constructions.

We  wrote four papers \cite{9,10,11,12} which contained the theory of transfer to the continuum by an almost analytical method which made calculable  spectra extending to very high continuum energies. We also managed to treat resonant and non resonant continuum states on the same ground.

A few years later, during a sabbatical in Orsay I realized that our method would be well suited to study breakup from the newly discovered halo nuclei. We wrote two papers \cite{13} and \cite{14}.
Ref. \cite{13} was the first to introduce an accurate description of neutron angular distributions following breakup, while Ref.\cite{14} showed that there were ways different from the eikonal model used by most other authors, to calculate breakup and that such methods could be useful to understand detailed aspects of momentum distributions, such as shapes \cite{14a}, widths, initial angular momentum, etc.

While in Orsay, working with Nicole Vinh Mau on Coulomb dissociation of $^{11}$Li
\cite{15}, I had developed an interest on two-neutron halo nuclei and Coulomb breakup and the problem of treating it at the same time as nuclear breakup. Jerome Margueron had just arrived at Pisa for a three month stage and we started a collaboration which led us to write Refs.\cite{16} and \cite{17}.
Ref.\cite{17} contains one of David's  very clever ideas, namely a way to regularise the all order series for Coulomb breakup in the eikonal approximation, by substituting the first order, divergent, term with the corresponding term calculated in time dependent perturbation theory.

During a visit at MSU, discussing with Carlos Bertulani, I started to be intrigued by the problem of the  existence of proton halo nuclei, which seem to show contradictory characteristics depending on the reaction used to study them. We wrote a simple, intuitive paper \cite{18} with Carlos, trying to compare the behavior of neutron vs. proton halos.
Along the same lines we have worked very recently with Alvaro Garc\'ia-Camacho to extend the method so as to treat all multipoles of the Coulomb potential development \cite{19} and finally to be able to calculate accurately proton breakup. This project is still in progress \cite{20}.

Meanwhile another french student, Guillaume Blanchon, has joined our group. He came to Pisa for a stage and later on was accepted by the local Galileo Galilei graduate school as a Ph.D student. We studied first $^{10}$Li produced by transfer to the continuum \cite{21} and we have now just finished a challenging paper
on projectile fragmentation and the treatment of neutron-core resonances due to the final  state interaction. We have applied the method to the description of experimental data of invariant mass spectra of the unbound nucleus
$^{13}$Be \cite{22}.

... Now we are close to leave the safe valley of stability and cross the nuclear dripline. Which will be next adventure of David, Angela and their fearless young collaborators?

\section{ My personal experience with David}

I have allowed myself a short time to talk and a little space to write because when I start talking about David it is always difficult for me to keep 'normal' (as Ricardo Broglia once noticed) and  other people with whom I have collaborated become jealous. 

In fact the only person that could be allowed to be jealous is Verena, David's wife. But I believe that she will not, because she knows that as a woman, mother and wife, I have learned from her as much as I have learned from David as a physicist. And this 'human' aspect is very important because one of the reasons for our long-standing collaboration is that it is based not only on Physics but also on the personal relationship. Both Verena and David have been my guide and teachers in all important moments of my life.

And this brings me to the title of this short talk. The more striking aspect of David'd activity is his capacity of being a teacher, a master, {\it un Maestro}. He is such in his personal attitude: when we discuss without our younger collaborators (presently Guillaume Blanchon  and Alvaro Garc\'ia-Camacho, Jerome Margueron previously), his constant concern is that we should explain to them what we have understood out of the discussion and make sure that they understand as well. When he comes to Pisa he spends most of his time with Alvaro and Guillaume. 

Once I asked Guillaume to check some notes that David had sent us and to compare to our own calculation. He answered : `I will do that but you know that he is always right and that in the end we always do as he suggests'.
Another time David was in Switzerland on vacation with his old computer but no book.  I was finishing a paper \cite{23} with George Bertsch and I asked David some advice on the use of the T-matrix formalism. For about a month we developed a correspondence which from David's side would easily account for a chapter of a Scattering Theory book. The amazing thing is that David wrote all of his notes
by heart without  consulting any book and he had all factors and constants right and ... the T-matrix formalism was in a mixed representation which is something you do not find in any standard textbook... But David is in fact a living textbook, a living treasure, as they would say in Japan.

Finally, I read recently an interesting definition of a leader: 'Leader is someone who leaves heirs'. Here you can see already three generations of David's heirs, because Massimo Di Toro opened my path to David and by an incredible coincidence yesterday it was the 25$^{th}$
anniversary since I graduated from Oxford (13$^{th}$ June 1981!) and already
Jerome, Alvaro and Guillaume are following from my side and there are many others all over the world.

So thanks to David from all of us and thanks to God for having given
him to us.

{\bf Acknowledgments}

Varenna Conference on Nuclear Reaction Mechanisms has been for the last thirty years one of the most successful  meetings in our field. The merit goes to Ettore Gadioli with his meticulous organization. In recent times each edition has been dedicated to an outstanding Nuclear Physicist. I wish to thank Ettore for his choice of dedicating the 2006 edition to David Brink.

\begin{figure}[h]\center
              \caption{Oxford, 13$^{th}$ June 1981: with Massimo Di Toro the day of my graduation. }                  
  \label{fig1}     \end{figure}

\begin{figure}[h]
                     \caption{Pisa, Summer 2002: discussing phase shift behavior with Guillaume Blanchon and having coffee on the Lungarno with Jerome Margueron.  }                  
  \label{fig2}     \end{figure}

\begin{figure}[h]\center
              \caption{Fauglia, July 2005: David's 75$^{th}$ birthday party. From left to right: Alex Babcenco, Silvia Lenzi and  Daniel Napoli, David, Ken Konishi, Alvaro, Verena, Guillaume, Ignazio Bombaci and his girlfriend Giusi, Hans-Peter Pavel, Lucia Vitturi. Front row: Ica Stancu and Angela.}                  
  \label{fig3}     \end{figure}

\end{document}